\documentstyle[12pt]{article}
\begin{document}
\begin{center}
{\bf THE NONCOMMUTATIVE SPECIAL RELATIVITY}
\vskip2truecm
M. Lagraa\\
\vskip1truecm
Laboratoire de Physique Th\'eorique\\
Universit\'e d'Oran Es-S\'enia, 31100, Alg\'erie\\
and\\
Laboratoire de Physique Th\'eorique et Hautes \'energies\footnote{Laboratoire
associ\'e au Centre National de la Recherche Scientifique - URA D0063},
Batiment 211,\\Universit\'e Paris XI, 91405 ORSAY,France\\
\vskip1truecm
\end{center}
{\bf Abstract :} We adapt the axioms of the quantum mechanics to the quantum
Minkowski space-time coordinates and their transformations under the quantum
Lorentz group to show how we can formulate the noncommutative special
relativity and its quantum physical observables. we establish in this
formalism the quantum analog of the lifetime dilatation formula and the
relativistic relations between the energy-momentum four-vector and the mass
and the velocity. From the explicit construction of states, we
establish the causality principle in the noncommutative special relativity and
show that for a free particle moving in the quantum Minkowski space-time, only
the length of the velocity and one of its components can be measured exactly
and simultaneously. In addition these observables present discret spectrums which
imply quantized lifetimes of moving unstable particles.\\
PACS numbers: 03.65.Fd, 03.30.+p, 03.65.w, 03.65.Ca
\newpage
\section{Introduction}
It is well known that a given physical theory describes well only a limited
class of phenomena measured in certain range of precision and desagrees with
other phenomena beyond this range which need a more general theory for their
description. In general, the latter involves new fondamental constant which
can be viewed as a kind of deformation parameter. The old theory can be
recovered in the limit where the new fundamental constant vanishes ( which
in general means that we do not reach enough precision to detect the effects
tied to this fundamental constant). The special relativity and the quantum
mechanics provide most famous examples of such a theory generalization. The
special relativity may be regarded as a deformed theory of Galilei's
relativity where the deformation parameter is the inverse of the velocity of
light and the quantum mechanics is a deformed theory of the classical
mechanics with the Planck constant as deformation parameter.\\
On the other hand, in view of the enormous and unsuccessful effort to overcome
the divergence difficulties in the relativistic quantum field theories, it is
believed that the framework of special relativity has to be changed.\\
In the past few years, attention has been paid to formulate the particle
evolution in quantum Minkowski space through the construction of the q-analogs
of the relativistic plane waves [1] or the Hilbert space representation of a
q-deformed Minkowski space [2]. Despite all the theoretical interests, the
relevance of the noncommutative Minkowski space-time and its transformations
under the quantum Lorentz group to measurable effects in particle physics has
not been discussed very much. \\
The above considerations make especially interesting the study of the
noncommutative special relativity in the frame of the quantum Minkowski
space-time and its transformations under the quantum Lorentz group to derive
measurable observables describing the evolution of particles in the
noncommutative space-time.\\
In this paper, we adapt the quantum mechanics axioms to the quantum
coordinates of Minkowski space-time and their transformations under quantum
Lorentz group to show how one can derive the quantum physical observables
corresponding to those of the usual special relativity.\\
This paper is organized as follows: In section 2, we recall the basic
properties derived from the correspondence between quantum $SL(2,C)$ and
Lorentz groups developped in [3]. In section 3, we consider Hilbert spaces on
which the generators of the Hopf algebras corresponding to the quantum Lorentz
symmetries and the quantum Minkowski space-time act. We establish the
transformation rules of states according to the properties of coordinate
transformations which are given in terms of tensorial products. The structure
of the commutation rules between the generators of the Lorentz symmetries and
the coordinates of the Minkowski space-time permits us to introduce the
velocity operator and to study the boost from which we establish the
quantum analog of the lifetime dilatation formula of unstable particles. We
end this section by nothing that this construction is also valid if we replace
the coosrdinates $X_{N}$ of the quantum Minkowski space-time by the
energy-momentum four-vector $P_{N}$. In this case we obtain the analog of the
usual relativistic formulas giving the energy and the vector-momentum in terms
of the mass and the velocity. In section 4, we carry on the investigation of
the Hilbert space states to describe the evolution of particles in the quantum
Minkowski space-time. From this state investigation, we establish the
principle of the causality in the noncommutative special relativity and show
that for a free particle moving in the noncommutative space-time, only the
length of the velocity and one of its components can be measured exactly. In
addition these observables present discret spectrums which lead to a quantized
lifetime of unstable paricles. We conclude this paper in section 5 by
discussing the case of particles moving in the light-cone. We also derive
directly from this formalism construction of the quantum Lorentz subgroup of
the three dimensional rotation of the space coordinates leaving the time
coordinate invariant. This permits us to deduce the properties of quantum
spheres.
\section{The quantum Lorentz group}
To explore the different properties of the quantum Lorentz group and the
quantum Minkowski space time, it is very convenient to represent the
generators $\Lambda_{N}^{~M}~(N,~M=~0,~1,~2,~3)$ of the quantum Lorentz group
in terms of generators $M_{\alpha}^{~\beta}~(\alpha, \beta=1,2)$ and
$(M_{\alpha}^{~\beta})^{\star} =M_{\dot{\alpha}}^{~\dot{\beta}}$ of the
quantum $SL(2,C)$ group which is well known. Let us start by recalling in this
section, some basic properties derived from the correspondence between
quantum $SL(2,C)$ and Lorentz groups developped in [3].\\
The unimodularity of $M_{\alpha}^{~\beta}$ is expressed by
$\varepsilon_{\alpha\beta}M_{\gamma}^{~\alpha} M_{\delta}^{~\beta}=
\varepsilon_{\gamma\delta}I_{\cal A}$,
$\varepsilon^{\gamma\delta}M_{\gamma}^{~\alpha} M_{\delta}^{~\beta}=
\varepsilon^{\alpha\beta}I_{\cal A}$, and
$\varepsilon_{\dot{\alpha}\dot{\beta}}M_{\dot{\gamma}}^{~\dot{\alpha}}
M_{\dot{\delta}}^{~\dot{\beta}}=
\varepsilon_{\dot{\gamma}\dot{\delta}}I_{\cal A}$,
$\varepsilon^{\dot{\gamma}\dot{\delta}}M_{\dot{\gamma}}^{~\dot{\alpha}}
M_{\dot{\delta}}^{~\dot{\beta}}=
\varepsilon^{\dot{\alpha}\dot{\beta}}I_{\cal A}$ where $I_{\cal A}$ is the
unity of the $\star$ algebra ${\cal A}$ generated by $M_{\alpha}^{~\beta}$ and the
spinor metric $\varepsilon_{\alpha\beta}$ and its inverse
$\varepsilon^{\alpha\beta}~(\varepsilon_{\alpha\delta}
\varepsilon^{\delta\beta}=\delta_{\alpha}^{\beta}=
\varepsilon^{\beta\delta}\varepsilon_{\delta\alpha})$ satisfy
$(\varepsilon_{\alpha\beta})^{\star} =\varepsilon_{\dot{\beta}\dot{\alpha}}$
and $(\varepsilon^{\alpha\beta})^{\star}=\varepsilon^{\dot{\beta}\dot{\alpha}}$.
The commutation rules are given by
$M_{\alpha}^{~\gamma}(f_{\pm\gamma}^{~~\beta}\star a) =
(a \star f_{\pm\alpha}^{~~\gamma})M_{\gamma}^{~\beta}$ and
$M_{\dot{\alpha}}^{~\dot{\gamma}}(f_{\pm\dot{\gamma}}^{~~\dot{\beta}}\star a)=
(a \star f_{\pm\dot{\alpha}}^{~~\dot{\gamma}})M_{\dot{\gamma}}^{~\dot{\beta}}$
, for any $a\in \cal A$, where $f_{\pm\alpha}^{~~\beta}$ and
$f_{\pm\dot{\alpha}}^{~~\dot{\beta}}$ $\in \cal A'$, dual to the Hopf algebra
$\cal A$, are linear functionals $:\cal A \rightarrow  \cal C$. The
convolution product is defined as $f\star a=(id \otimes f)\Delta(a)$ and
$a\star f=(f\otimes id)\Delta(a)$. The subalgebra
${\cal A}'_{0} \subset {\cal A}'$ generated by $f_{\pm\alpha}^{~~\beta}$ and
$f_{\pm\dot{\alpha}}^{~~\dot{\beta}}$ is a Hopf algebra acting on the
genrators of $\cal A$ as $f_{\pm\alpha}^{~~\beta}(M_{\gamma}^{~\delta})=
a^{\mp \frac{1}{2}}R^{\pm\delta\beta}_{~\alpha\gamma}$ and
$f_{\pm\dot{\alpha}}^{~~\dot{\beta}}(M_{\dot{\gamma}}^{~\dot{\delta}})=
a^{\mp \frac{1}{2}}
R^{\pm\dot{\delta}\dot{\beta}}_{~\dot{\alpha}\dot{\gamma}}$ [4] where
$a\not=0$ is a real number satisfying
$\varepsilon_{\alpha\beta}\varepsilon^{\alpha\beta}
= -(a + a^{-1})=-Q$. The R-matrices $R^{\pm\delta\beta}_{~\alpha\gamma}=
\delta^{\delta}_{\alpha}\delta^{\beta}_{\gamma} + a^{\pm 1}
\varepsilon^{\delta\beta}\varepsilon_{\alpha\gamma}~(
R^{\pm\dot{\delta}\dot{\beta}}_{~\dot{\alpha}\dot{\gamma}}=
\delta^{\dot{\delta}}_{\dot{\alpha}}\delta^{\dot{\beta}}_{\dot{\gamma}} +
a^{\pm 1}\varepsilon^{\dot{\delta}\dot{\beta}}
\varepsilon_{\dot{\alpha}\dot{\gamma}})$ satisfy
$R^{\pm\delta\beta}_{~\alpha\gamma}R^{\mp\rho\sigma}_{~\delta\beta}=
\delta^{\rho}_{\alpha}\delta^{\sigma}_{\gamma}~(
R^{\pm\dot{\delta}\dot{\beta}}_{~\dot{\alpha}\dot{\gamma}}
R^{\mp\dot{\rho}\dot{\sigma}}_{~\dot{\delta}\dot{\beta}}=
\delta^{\dot{\rho}}_{\dot{\alpha}}\delta^{\dot{\sigma}}_{\dot{\gamma}})$, the
Hecke conditions $(R^{\pm}+a^{\pm 2})(R^{\pm}-1)=0$ and the Yang-Baxter
equations [5].\\
Besides commutation rules between undotted or dotted generators only, we need
to specify commutation rules between undotted and dotted generators which are
controlled by $R$-matrices
$f_{\pm\alpha}^{~\beta}(M_{\dot{\gamma}}^{~\dot{\delta}})=
R^{\pm\dot{\delta}\beta}_{~\alpha\dot{\gamma}}$ which we assume they take the
form of the R matrix of the quantum $SU(2)$ group. More precisely,
the generators $M_{\alpha}^{~\beta}$ must satisfy the unitarity condition,
$M_{\dot{\alpha}}^{~\dot{\beta}}=S(M_{\beta}^{~\alpha})$ and
$M_{\alpha}^{~\beta}=S^{-1}(M_{\dot{\beta}}^{~\dot{\alpha}})$, when
they are considered as belonging to a range of functionals
$f_{\pm\alpha}^{~~\beta}$,
$f_{\pm\alpha}^{~~\beta}(M_{\dot{\gamma}}^{~\dot{\delta}})=
R^{\pm\dot{\delta}\beta}_{~\alpha\dot{\gamma}}=
f_{\pm\alpha}^{~~\beta}(S(M_{\delta}^{~\gamma}))=a^{\pm\frac{1}{2}}
R^{\mp\beta\gamma}_{~\delta\alpha}$ and
$f_{\pm\alpha}^{~~\beta}(S(M_{\dot{\gamma}}^{~\dot{\delta}}))=
R^{\mp\beta\dot{\delta}}_{~\dot{\gamma}\alpha}=
\varepsilon_{\dot{\gamma}\dot{\rho}}f_{\pm\alpha}^{~~\beta}(M_{\dot{\sigma}}
^{~\dot{\rho}})\varepsilon^{\dot{\sigma}\dot{\delta}}$. In this case,
the spinor metrics must satisfy an additional condition,
$\varepsilon_{\alpha\beta}=-\varepsilon^{\beta\alpha}$ and
$\varepsilon_{\dot{\alpha}\dot{\beta}}=-
\varepsilon^{\dot{\beta}\dot{\alpha}}$, required by consistency condition between
the unitarity and the unimodularity of the quantum $SU(2)$ group generators.
The functionals $f_{\pm\alpha}^{~~\beta}$ and
$f_{\pm\dot{\alpha}}^{~~\dot{\beta}}$ satisfy the conditions
$f_{\pm\dot{\alpha}}^{~~\dot{\beta}} =\tilde{f}_{\pm\beta}^{~~\alpha}$ and
$\tilde{f}_{\pm\dot{\alpha}}^{~~\dot{\beta}} = f_{\pm\beta}^{~~\alpha}$ where
$f_{\pm\beta}^{~~\alpha}=\tilde{f}_{\pm\beta}^{~~\alpha}\circ S =
\varepsilon^{\alpha\delta}\tilde{f}_{\pm\delta}^{~~\gamma}
\varepsilon_{\gamma\beta}$
and
$f_{\pm\dot{\alpha}}^{~~\dot{\beta}}=
\tilde{f}_{\pm\dot{\alpha}}^{~~\dot{\beta}}\circ S^{-1}=
\varepsilon_{\dot{\alpha}\dot{\gamma}}
\tilde{f}_{\pm\dot{\delta}}^{~~\dot{\gamma}}
\varepsilon^{\dot{\delta}\dot{\beta}}$. These functionals
also satisfy $(f_{\pm\beta}^{~~\alpha}(a))^{\star} =
f_{\mp\dot{\alpha}}^{~~\dot{\beta}}(S(a^{\star}))$ and
$(f_{\pm\dot{\beta}}^{~~\dot{\alpha}}(a))^{\star}
= f_{\mp\alpha}^{~~\beta}(S(a^{\star}))$.\\
With these hypothesis on the commutation rules, it is shown in [3] that
the generators $\Lambda_{N}^{~M}$ of quantum Lorentz group may
be written in terms of those of quantum $SL(2,C)$ group as 
$\Lambda_{N}^{~M} =\frac{1}{Q}\varepsilon_{\dot{\gamma}\dot{\delta}}
\overline{\sigma}_{N}^{~\dot{\delta}\alpha}M_{\alpha}^{~\sigma}
\sigma^{M}_{~\sigma\dot{\rho}}M_{\dot{\beta}}^{~\dot{\rho}}
\varepsilon^{\dot{\gamma}\dot{\beta}}$. They are real,
$(\Lambda_{N}^{~M})^{\star}=\Lambda_{N}^{~M}$, and generate a Hopf algebra
$\cal L$ endowed with a coaction $\Delta$, a counit $\varepsilon$ and an
antipode $S$ acting as $\Delta(\Lambda_{N}^{~M}) =
\Lambda_{N}^{~K}\otimes \Lambda_{K}^{~M}$, $\varepsilon(\Lambda_{N}^{~M})=
\delta_{N}^{M}$ and $S(\Lambda_{N}^{~M})=
G_{\pm NK}\Lambda_{L}^{~K}G_{\pm}^{LM}$ respectively. $G_{\pm}^{NM}$ is an
invertible and hermitian quantum metric. It may be expressed in
terms of the four matrices $\sigma^{N}_{\alpha\dot{\beta}}~(N=0,1,2,3)$ as:
\begin{eqnarray*}
G_{\pm}^{~IJ} =\frac{1}{Q}Tr(\sigma^{I} \overline{\sigma}_{\pm}^{J}) =
\frac{1}{Q}\varepsilon^{\alpha \nu}\sigma^{I}_{~\alpha\dot{\beta}}
\overline{\sigma}_{\pm}^{J\dot{\beta}\gamma}\varepsilon_{\gamma\nu}=
\frac{1}{Q}Tr(\overline{\sigma}_{\pm}^{I} \sigma^{J})=
\frac{1}{Q} \varepsilon_{\dot{\nu}\dot{\gamma}}
\overline{\sigma}_{\pm}^{I\dot{\gamma}\alpha}
\sigma^{J}_{~\alpha\dot{\beta}} \varepsilon^{\dot{\nu}\dot{\beta}}
\end{eqnarray*}
where $\sigma^{n}_{\alpha\dot{\beta}}~(n=1,2,3)$ are the usual Pauli matrices,
$\sigma^{0}_{\alpha\dot{\beta}}$ is the identity matrix and
$\overline{\sigma}_{\pm}^{I\dot{\alpha} \beta} = \varepsilon^{\dot{\alpha}
\dot{\lambda}}R^{\mp \sigma \dot{\rho}}_{~~\dot{\lambda}\nu}
\varepsilon^{\nu \beta}\sigma^{I}_{~\sigma \dot{\rho}}$. The inverse of the
Minkowskian metric may be written under the form
$G_{\pm IJ}=\frac{1}{Q}Tr(\overline{\sigma}_{J}\sigma_{\pm I})=
\frac{1}{Q} \varepsilon_{\dot{\nu}\dot{\gamma}}
\overline{\sigma}_{J}^{~\dot{\gamma}\alpha}
\sigma_{\pm I\alpha\dot{\beta}} \varepsilon^{\dot{\nu}\dot{\beta}}$ where
$\sigma_{\pm I\alpha \dot{\beta}} = G_{\pm IJ}\sigma^{J}_{\alpha\dot{\beta}}$.
The form of the antipode of $\Lambda_{N}^{~M}$ guarantees the orthogonality
condition on the generators of quantum Lorentz group as:
\begin{eqnarray}
G_{\pm NM}\Lambda_{L}^{~N}\Lambda_{K}^{~M} = G_{\pm LK}I_{\cal L}~~and~~
G_{\pm}^{~LK}\Lambda_{L}^{~N}\Lambda_{K}^{~M} = G_{\pm}^{~NM}I_{\cal L}
\end{eqnarray}
where $I_{{\cal L}}$ is the unity of the algebra $\cal L$.\\
The completeness relations are given by $\sigma^{I~\dot{\beta}}_{\alpha}
\overline{\sigma}_{I\dot{\rho}}^{~~~\sigma}=
Q\delta^{\sigma}_{\alpha}\delta_{\dot{\rho}}^{\dot{\beta}}$ and
$\sigma_{\pm I~\dot{\beta}}^{~~\alpha}
\overline{\sigma}^{I\dot{\rho}}_{\pm~\sigma}=
Q\delta_{\sigma}^{\alpha}\delta^{\dot{\rho}}_{\dot{\beta}}$ where the
undotted and dotted spinorial indices are raised and lowered as
$\sigma^{I\alpha}_{~~\dot{\beta}} = \sigma^{I}_{~\rho\dot{\beta}}
\varepsilon^{\rho\alpha}$ and $\sigma^{I~\dot{\beta}}_{~\alpha}=
\varepsilon^{\dot{\beta}\dot{\rho}}\sigma^{I}_{\alpha\dot{\rho}}$. These
completeness relations may be used to convert a vector to a bispinor and vice
versa
\begin{eqnarray*}
X_{\alpha\dot{\beta}} = X_{I}\sigma^{I}_{\alpha\dot{\beta}}\Leftrightarrow 
X_{I}= \frac{1}{Q}\varepsilon^{\alpha\nu}X_{\alpha\dot{\beta}}
\overline{\sigma}_{\pm}^{J\dot{\beta}\delta}\varepsilon_{\delta\nu}
G_{\pm JI}~~or ~~
X_{I}= \frac{1}{Q}\varepsilon_{\dot{\nu}\dot{\beta}}
\overline{\sigma}_{I}^{~\dot{\beta}\alpha}X_{\alpha\dot{\delta}}
\varepsilon^{\dot{\nu}\dot{\delta}}
\end{eqnarray*}
where $X_{I}$ are real elements of the right invariant basis of a bimodule
$\cal M$ over the algebra $\cal L$ which transform under the left coaction
$\Delta_{L}:{\cal M} \rightarrow {\cal L}\otimes {\cal M}$ as
\begin{eqnarray}
\Delta_{L}(X_{I})=\Lambda_{I}^{~K}\otimes X_{K}.
\end{eqnarray}
The functional $f_{\alpha}^{~\beta}$ of $SL(2,C)$ induce that of Lorentz
group $F_{\pm L}^{~~K}: {\cal L} \rightarrow C$ given by $F_{\pm L}^{~~K} =
\frac{1}{Q}(\tilde{f}_{\mp \dot{\beta}}^{~~~\dot{\alpha}}
\overline{\sigma}_{L\dot{\alpha}}^{~~~\delta} \star f_{\pm\delta}^{~~~\gamma}
\sigma^{K~\dot{\beta}}_{~\gamma})$. They controle the noncommutatitivity
between elements of $\cal L$ and $X_{I}$ as
\begin{eqnarray}
\Lambda_{L}^{~I}(F_{\pm I}^{~~K} \star a) &=&
(a \star F_{\pm L}^{~~I})\Lambda_{I}^{~K} ~,\\
X_{\pm L}a=(a \star F_{\pm L}^{~~K})X_{\pm K}~&and&~X_{(a)L}X_{(b)K}=
F_{(b)K}^{~~~~N}(S(\Lambda_{L}^{~M}))X_{(b)N}X_{(a)M}
\end{eqnarray}
where $a\in \cal L$ and the indices $a,b=\pm$. These functionals satisfy
$F_{\pm L}^{~~K}(\delta_{N}^{M})=\delta_{L}^{K}\delta_{N}^{M}$,
$(F_{\pm L}^{~~K}(a))^{\star}=F_{\pm L}^{~~K}(S(a^{\star}))$ for any
$a\in \cal L$ and the relations
$G_{\pm}^{~MN}F_{\pm N}^{~~L} \star F_{\pm M}^{~~K}(a) = G_{\pm}^{~KL}
\varepsilon(a)$ and $G_{\pm KL}F_{\pm N}^{~~L} \star F_{\pm M}^{~~K}(a) =
G_{\pm MN}\varepsilon(a)$ which imply that the length of the quantum
four-vector, $G^{LK}X_{L}X_{K}$, is bi-invariant, central and real. Since it
commutes with everything, it is of the form $G^{LK}X_{L}X_{K}=-\tau^{2}
I_{\cal L}$ where $\tau^{2}$ is a real number.\\
We can also show that the quantum symmetrization of the Minkowski metric is
given by ${\cal R}^{\pm NM}_{~KL}G_{\pm}^{KL} = G_{\pm}^{NM}$ and
${\cal R}^{\pm NM}_{~KL}G_{\pm NM} = G_{\pm KL}$ where
${\cal R}^{\pm NM}_{~KL} = F_{\pm K}^{~~M}(\Lambda_{L}^{~N})$ satisfy
the Yang-Baxter equations and the cubic Hecke conditions
$({\cal R}^{\pm} + a^{\pm2})({\cal R}^{\pm} + a^{\mp2})
({\cal R}^{\pm} - 1)=0$.\\
In following we shall consider the right invariant basis $X_{I}=X_{+I}$ of
the bimodule algebra ${\cal M}$as a quantum coordinate system of the Minkowski
space-time $\cal M$ provided with
the metric $G^{IJ}=G_{+}^{~IJ}$. $X_{0}$ represents the time operator and
$X_{i}~(i=1,2,3)$ represent the space coordinate operators.\\
To make the explicit calculation of the different commutation rules, we take
an adequate choice of Pauli hermitian matrices of the form
\begin{eqnarray*}
\sigma^{0}_{\alpha \dot{\beta}} =\left(\begin{array}{cc}
1 & 0\\
0 & 1
\end{array}
\right)~~,~~
\sigma^{1}_{\alpha \dot{\beta}} =\left(\begin{array}{cc}
0 & 1\\1 & 0
\end{array}
\right)~~,~~
\sigma^{2}_{\alpha \dot{\beta}} =\left(\begin{array}{cc}
0 & -i\\
i &  0
\end{array}
\right)~~,~~
\sigma^{3}_{\alpha \dot{\beta}} =\left(\begin{array}{cc}
q & 0\\
0 &-q^{-1}
\end{array}
\right)
\end{eqnarray*}
and the spinorial metrics 
$\varepsilon_{\alpha\beta}=-\varepsilon^{\alpha\beta}=
\varepsilon_{\dot{\beta}\dot{\alpha}}=-\varepsilon^{\dot{\beta}\dot{\alpha}}
=\left(\begin{array}{cc}
0& -q^{-\frac{1}{2}}\\q^{\frac{1}{2}}&0 
\end{array}
\right)$ which presents an advantage arising from the fact that, in this
representation, we have $\overline{\sigma}_{0}^{~\dot{\alpha}\beta} =
-\sigma^{0}_{\alpha \dot{\beta}}=-\delta_{\alpha}^{\beta}$,
$\overline{\sigma}_{N\dot{\alpha}\alpha}=\overline{\sigma}_{N\dot{1}1} +
\overline{\sigma}_{N\dot{2}2}=-(q+q^{-1})\delta_{N}^{0}=-Q\delta_{N}^{0} $
and $\sigma^{N\alpha\dot{\alpha}}=Q\delta_{0}^{N}$. We shall see in the next
section that these properties lead directly to the restriction of the quantum
lorentz group to the quantum subgroup of the three dimensional space rotations
by restricting the quantum $SL(2,C)$ group to the $SU(2)$ group. These
representations also give a form of the quantum metric $G^{NM}$ into two
independent blocks, one for the time component $X_{0}$ and the other for
space components $X_{k}~(k=1,2,3)$. The non vanishing elements of the metric
are $G^{00}=-q^{-\frac{3}{2}}$, $G^{11}=G^{22}=G^{33}=q^{\frac{1}{2}}$,
$G^{12}=-G^{21}=-iq^{\frac{1}{2}}\frac{q-q^{-1}}{Q}$ and the non vanishing
elements of its inverse are $G_{00}=-q^{\frac{3}{2}}$,
$G_{11}=G_{22}=q^{-\frac{1}{2}}\frac{Q^{2}}{4}$, $G_{33}=q^{-\frac{1}{2}}$,
$G_{12}=-G_{21}=iq^{-\frac{1}{2}}\frac{(q-q^{-1})Q}{4}$. In the classical
limit $q=1$, this metric reduces to the classical Minkowski metric with
signature $(-,+,+,+)$.\\
\section{\bf The lifetime dilatation in the quantum Minkowski space-time}
Before to show how to derive $q$-analog physical properties of the usual
special relativity, let us investigate the properties of the $\cal R$ matrix
and its consequences on the different commutation rules. Let us recall that
as a range of the functionals $f_{\alpha}^{~\beta}$ the generators
$M_{\alpha}^{~\beta}$ of the Hopf algebra $\cal A$ satisfy unitarity
condition. Hence as a range of the functionals $F_{N}^{~M}$, the generators of
$\cal L$ have the following properties
\begin{eqnarray}
\Lambda_{N}^{~0} =\frac{1}{Q}
\overline{\sigma}_{N\dot{\gamma}}^{~~~\alpha}M_{\alpha}^{~\sigma}
\sigma^{0}_{~\sigma\dot{\rho}}S(M_{\rho}^{~\beta})
\varepsilon^{\dot{\gamma}\dot{\beta}} &=&\frac{1}{Q}
\overline{\sigma}_{N\dot{\gamma}}^{~~~\alpha}
\varepsilon^{\dot{\gamma}\dot{\alpha}}\nonumber\\
=-\frac{1}{Q}\overline{\sigma}_{N\dot{\gamma}\gamma}&=& \delta_{N}^{0}
\end{eqnarray}
and
\begin{eqnarray}
\Lambda_{0}^{~M} =\frac{1}{Q}\varepsilon_{\dot{\gamma}\dot{\delta}}
\overline{\sigma}_{0}^{~\dot{\delta}\alpha}M_{\alpha}^{~\sigma}
\sigma^{M}_{~\sigma\dot{\rho}}S(M_{\rho}^{~\beta})
\varepsilon^{\dot{\gamma}\dot{\beta}} &=& -\frac{1}{Q}\varepsilon^{\alpha\gamma}
M_{\alpha}^{~\sigma}\sigma^{M}_{~\sigma\dot{\rho}}\varepsilon_{\rho\delta}
M_{\gamma}^{~\delta}=\nonumber\\
\frac{1}{Q} \varepsilon^{\dot{\delta}\dot{\rho}}
\sigma^{M}_{~\sigma\dot{\rho}}\varepsilon^{\sigma\delta} &=& \frac{1}{Q}
\sigma^{M\delta\dot{\delta}}=\delta_{0}^{M}
\end{eqnarray}
from which, we get
\begin{eqnarray}
F_{N}^{~M}(\Lambda_{P}^{~0}) =\delta_{N}^{M}\delta_{P}^{0} =
{\cal R}^{0M}_{NP},~and~F_{N}^{~M}(\Lambda_{0}^{~Q}) =
\delta_{N}^{M}\delta_{0}^{Q}={\cal R}^{QM}_{N0}.
\end{eqnarray}
The quantum symmetrization of the coordinates may be written, from the right
relation of (4), as
\begin{eqnarray}
X_{L}X_{K} = {\cal R}^{NM}_{LK}X_{N}X_{M}
\end{eqnarray}
leading for $K=0$ to
\begin{eqnarray}
X_{L}X_{0} - {\cal R}^{NM}_{L0}X_{N}X_{M}=X_{0}X_{L}
\end{eqnarray}
which shows that the time coordinate operator commutes with the space
coordinate operator. After a straighforward computation (8) gives the
commutation relations between space coordinates as:
\begin{eqnarray}
X_{3}Z - q^{2}ZX_{3}=(q - q^{-1})X_{0}Z,\\
X_{3}\overline{Z} - q^{-2}\overline{Z}X_{3}=
- q^{-2}(q - q^{-1})X_{0}\overline{Z},\\
Z\overline{Z}-\overline{Z}Z = (q^{2} - q^{-2})X_{3}^{2} +
q^{-1}(q^{2} - q^{-2})X_{0}X_{3}
\end{eqnarray}
where $Z=X_{1}+iX_{2}$ and $\overline{Z}=X_{1}-iX_{2}$. The four-vector
length $G^{IJ}X_{I}X_{J} = -\tau^{2}I$ may be written of the form
\begin{eqnarray}
q^{-\frac{3}{2}}X_{0}^{2} - \frac{q^{\frac{3}{2}}}{Q}Z\overline{Z} -
\frac{q^{-\frac{1}{2}}}{Q}\overline{Z}Z - q^{\frac{1}{2}}X_{3}^{2} = \tau^{2}.
\end{eqnarray}
Note that by redefining the Minkowski space-time coordinates as $C=qX_{0}-X_{3}$,
$D=q^{-1}X_{0}+X_{3}$, $A=Z$ and $B=\overline{Z}$ we recover the commutation
relations and the four-vector length given in Ref.[2,6,7].\\
As a consequence of (7), the relation (3) gives, for $a=\Lambda_{N}^{~0}$,
\begin{eqnarray}
\Lambda_{L}^{~I}\Lambda_{N}^{~P}F_{I}^{~K}(\Lambda_{P}^{~0})=
\Lambda_{L}^{~K}\Lambda_{N}^{~0}=F_{L}^{~I}(\Lambda_{N}^{~P})
\Lambda_{P}^{~0}\Lambda_{I}^{~K}={\cal R}^{PI}_{LN}\Lambda_{P}^{~0}
\Lambda_{I}^{~K}
\end{eqnarray}
which reduces to 
\begin{eqnarray}
\Lambda_{L}^{~0}\Lambda_{N}^{~0}={\cal R}^{PI}_{LN}\Lambda_{P}^{~0}
\Lambda_{I}^{~0}
\end{eqnarray}
for $K=0$ and to
\begin{eqnarray}
\Lambda_{L}^{~K}\Lambda_{0}^{~0}=\Lambda_{0}^{~0}\Lambda_{L}^{~K}
\end{eqnarray}
for $N=0$. The commutation relations between the Minkowski space-time
coordinates $X_{L}$ and the generators of the quantum Lorentz group
$\Lambda_{N}^{~M}$ are given by the left relation of (4) which gives for
$a=\Lambda_{N}^{~0}$
\begin{eqnarray}
X_{L}\Lambda_{N}^{~0} = F_{L}^{~K}(\Lambda_{N}^{~P})\Lambda_{P}^{~0}
X_{K} = {\cal R}^{PK}_{LN}\Lambda_{P}^{~0}X_{K}.
\end{eqnarray}
This relation reduces to
\begin{eqnarray}
X_{L}\Lambda_{0}^{~0} = F_{L}^{~K}(\Lambda_{0}^{~P})\Lambda_{P}^{~0}
X_{K} = \Lambda_{0}^{~0}X_{L}
\end{eqnarray}
for $N=0$.\\
We are now ready to show that the above commutation relations suffice to
construct the physical states and the different $q$-deformed observables
derived from the quantum boost of particles at rest. Since the coordinates
and their transformations are operators, we consider, for the usual quantum
mechanics, that the evolution of a free particle in the noncommutative
special relativity is described by Hilbert space states which are common
eingenstates of a set of commuting elements of the quantum algebras
${\cal M}$ and ${\cal L}$. The corresponding eingenvalues give the measurable
quantities tied to the evolution of this particle.\\
First we see from the relations (16) and (18) that $\Lambda_{0}^{~0}$
commutes with the algebras $\cal L$ and $\cal M$, then it
is a real c-number. Therefore, as in textbooks of special relativity we set
$\Lambda_{0}^{~0}=\gamma I_{\cal L} $ and $\Lambda_{I}^{~0}=\gamma V_{I}$
where $V_{0}=I_{\cal L}$ and $V_{i}$ are the components of the velocity
operator. Due to the fact that $\gamma$ is a real c-number and
$\Lambda_{I}^{~0}$ are real operators, the components $V_{i}$ of the velocity
are real operators. From (15) we see that $\Lambda_{I}^{~0}$ fulfil the same
commutation relations (8) of the coordinates $X_{I}$, then $V_{I}$ satisfy
the commutations rules $V_{L}V_{K}=R_{LK}^{NM}V_{N}V_{M}$ which give the same
relations (10-12) by replacing $X_{i}$ by $V_{i}$ and $X_{0}$ by $1$. With
these notations, the orthogonality relations (1) give
\begin{eqnarray}
G^{IJ}\Lambda_{I}^{~0}\Lambda_{J}^{~0} =
\gamma^{2}(G^{00} + G^{ij}V_{i}V_{j})= G^{00}
\end{eqnarray}
leading to
\begin{eqnarray}
\gamma =(1 - |v|_{q}^{2})^{-\frac{1}{2}}.
\end{eqnarray}
Since $\gamma$ is a real c-number, the length of the velocity,
$-\frac{G^{ij}}{G^{00}}V_{i}V_{j} = |v|_{q}^{2}$, is also a
real c-number. Then both $\gamma$ and $|V|_{q}$ can be measured exactly.
$\Lambda_{0}^{~0}=(\varepsilon^{\alpha\delta}M_{\alpha}^{~\sigma})
(\varepsilon^{\alpha\delta}M_{\alpha}^{~\sigma})^{\star}$ is a real and
positive operator, then $\gamma >0$. From (20) we see that the upper limit
of the velocity is $1$ (the light velocity). This point will be discussed in
more details in the next section.\\

Now, we show how the states transform under a Lorentz group. To do that,
we observe that we are in presence of two quantum algebras, $\cal L$
corresponds to the quantum symmetries and $\cal M$ is generated by the
quantum coordinates of the Minkowski space-time. Since the coordinates
transform through the left coaction (2) as a tensor product
${\cal L} \otimes {\cal M}$, we consider the space states of positive norm as
a tensor product ${\cal H}_{\cal L} \otimes {\cal H}_{\cal M}$ of two Hilbert
spaces in which the algebras $\cal L$ and $\cal M$ act respectively (space of
representations of the coordinate transformations). Then in this formalism we
have to construct bases for ${\cal H}_{\cal M}$ and ${\cal H}_{\cal L}$ which
are eingenstate of the set of commuting elements of algebras $\cal L$ and
$\cal M$ separately.\\
From the commutation rules of the coordinates (9-12), we see that the set of
commuting operators are the proper time $\tau^{2}$, the time component
$X_{0}$ and one of the space components, for example $X_{3}$. Then in a fixed
reference system, a particle is described by a state belonging to the Hilbert
space ${\cal H}_{\cal M}$ labeled by the eigenvalues $\tau^{2}$, $t$ and
$x_{3}$, $|{\cal P}\rangle=|t,x_{3},\tau^{2}\rangle$ eingenstate of $X_{0}$,
$X_{3}$ and $\tau^{2}$ respectively
\begin{eqnarray}
X_{0}|{\cal P}\rangle=t |{\cal P}\rangle~~,~~
X_{3}|{\cal P}\rangle= x_{3} |{\cal P}\rangle~~,~~
\tau^{2}|{\cal P}\rangle=\tau^{2}|{\cal P}\rangle.
\end{eqnarray}
This state describes the evolution of a particle seen by an observer $O$ in a
coordinate system $X_{N}$. For a second observer $O'$ connected with the
observer $O$ by a quantum Lorentz transformation, the coordinate system tied
to $O'$ is given by $X'_{N}=\Delta_{L}(X_{N})=\Lambda_{N}^{~M}\otimes X_{M}$
which fulfil the same commutation rules (9-12) as $X_{N}$. Then the state
$|{\cal P}'\rangle$, describing the same particle seen by the observer $O'$ is
also labeled by the eigenvalues of $\tau^{2}$, $X'_{0}$ and $X'_{3}$ as
$|{\cal P}'\rangle=|t',x'_{3},\tau^{2}\rangle \in {\cal H}_{\cal M}$
satisfying:
\begin{eqnarray}
X'_{0}|{\cal P}'\rangle=t' |{\cal P}'\rangle~~,~~
X'_{3}|{\cal P}'\rangle= x'_{3} |{\cal P}'\rangle~~,~~
\tau^{2}|{\cal P}'\rangle=\tau^{2}|{\cal P}'\rangle.
\end{eqnarray}
Since the coordinates transform with tensorial products as
$X'_{N}=\Lambda_{N}^{~M}\otimes X_{M}$ it is natural to suppose that the
Hilbert state $|{\cal P}\rangle$ transforms into $|{\cal P}'\rangle$ as
\begin{eqnarray}
|{\cal P}'\rangle=|sym_{q}\rangle \otimes |{\cal P}\rangle
\end{eqnarray}
where $|sym_{q}\rangle$ is a basis of the Hilbert space ${\cal H}_{\cal L}$
where the generators $\Lambda_{N}^{~M}$ of the quantum Lorentz group are
represented. Note that:\\
- The state transformation (23) are different from those of the usual quantum
mechanics where the states transform with the unitary transformations $U(G)$
of the classical group $G$ as $|\Psi '\rangle=U(G)|\Psi\rangle$.\\
- At this stage there is no relation between the eigenvalues of space
coordinates and those of Lorentz group as velocities. So (23) may be regarded
as the quantum analog of a point in the configuration space of the classical
mechanics.\\
Under the quantum Lorentz transformation, the coordinates $X'_{N}$ act on
$|{\cal P}'\rangle$ as:
\begin{eqnarray}
X'_{0}|{\cal P}'\rangle = 
(\Lambda_{0}^{~0}\otimes X_{0})|{\cal P}'\rangle +
(\Lambda_{0}^{~k}\otimes X_{k})|{\cal P}'\rangle=
\Lambda_{0}^{~0}|sym_{q}\rangle\otimes X_{0}|{\cal P}\rangle +
\Lambda_{0}^{~k}|sym_{q}\rangle\otimes X_{k}|{\cal P}\rangle,\\
X'_{i}|{\cal P}'\rangle =
(\Lambda_{i}^{~0}\otimes X_{0})|{\cal P}'\rangle +
(\Lambda_{i}^{~k}\otimes X_{k})|{\cal P}'\rangle=
\Lambda_{i}^{~0}|sym_{q}\rangle\otimes X_{0}|{\cal P}\rangle +
\Lambda_{i}^{~k}|sym_{q}\rangle\otimes X_{k}|{\cal P}\rangle.
\end{eqnarray}
In the following we consider the simplest case where a particle at rest is
boosted with a velocity $v$. Let $|{\cal P}_{0}\rangle =|t,0,\tau^{2}\rangle$
a state describing a particle at rest. This state satisfy beside
$X_{0}|t,0,\tau^{2}\rangle=t|t,0,\tau^{2}\rangle$,
$\tau^{2}|t,0,\tau^{2}\rangle=\tau^{2}|t,0,\tau^{2}\rangle$ and
$X_{3}|t,0,\tau^{2}\rangle=0|t,0,\tau^{2}\rangle$ two additionnal conditions
\begin{eqnarray}     
X_{1}|{\cal P}\rangle= 0 |{\cal P}\rangle~~,~~
X_{2}|{\cal P}\rangle= 0 |{\cal P}\rangle.
\end{eqnarray}
The latter relations are possible because of the homogeneous feature of the
commutation rules (10-12) and do not desagree with (13). In the next section,
we shall show that this rest state is unique. By using the properties (26) of
the rest state, the transformations (24-25) reduce to
\begin{eqnarray}
X'_{0}|{\cal P}'\rangle = t'|{\cal P}'\rangle =
\Lambda_{0}^{~0}|sym_{q}\rangle\otimes X_{0}|{\cal P}_{0}\rangle=
\gamma|sym_{q}\rangle\otimes t|{\cal P}_{0}\rangle=
\gamma t'|{\cal P}'\rangle,\\
X'_{i}|{\cal P}'\rangle =
\Lambda_{i}^{~0}|sym_{q}\rangle\otimes X_{0}|{\cal P}_{0}\rangle =
V_{i}\gamma|sym_{q}\rangle\otimes t|{\cal P}_{0}\rangle=
V_{i}\gamma t|{\cal P}'\rangle= V_{i}t'|{\cal P}'\rangle.
\end{eqnarray}
These transformations require the knowledg of the commutation rules between
$\Lambda_{N}^{~0}=V_{N}\gamma$ only, which can be deduced from (9-12). Then
as the coordinate system, the state describing the quantum boost are labeled
by the length $|v|_{q}^{2}$ or $\gamma$ and its component $v_{3}$. And,
therefore the transformed Hilbert space state may be written under the form
of tensor product of the rest state $|{\cal P}_{0}\rangle$ and the boost
state labelled by $v_{3}$ and the length of the
velocity $|v|_{q}^{2}$ eigenvalues of the common eigenstate
$|v_{3},|v|_{q}^{2}\rangle$ of $V_{3}$ and
$-\frac{G^{ij}}{G^{00}}V_{i}V_{j}$ as:
\begin{eqnarray}
|{\cal P}'\rangle =|v_{3},|v|_{q}^{2}\rangle \otimes |{\cal P}_{0}\rangle.
\end{eqnarray}
On this state, the transformation (27-28) lead to
\begin{eqnarray}
X'_{0}|{\cal P}'\rangle = t'|{\cal P}'\rangle =
(\gamma\otimes t)|{\cal P}'\rangle~~and~~
X'_{3}|{\cal P}'\rangle = x'_{3}|{\cal P}'\rangle =
(v_{3}\otimes\gamma t)|{\cal P}'\rangle= v_{3}t'|{\cal P}'\rangle
\end{eqnarray}
which show the usual relation $x'_{3}=v_{3}t$ between the coordinate $x_{3}$
and the component $v_{3}$ of the velocity and the time for a free particle
moving with the velocity $|v|_{q}$. Note that we have assumed $t\geq 0$,
since $\gamma\geq 0$ then $t'=\gamma t\geq 0$. On the other hand the
combination of the invariance of the proper time $\tau^{2}$ with (26-28) and
their conjugate ($\langle{\cal P}|X'_{i}=\langle{\cal P}|V_{i}t$) gives
\begin{eqnarray*}
\langle{\cal P}'|G^{IJ}X'_{I}X'_{J}|{\cal P}'\rangle =
\langle{\cal P}'|G^{00}(X'^{2}_{0} &+& \frac{G^{ij}}{G^{00}}X'_{i}X'_{j})
|{\cal P}'\rangle =
\langle{\cal P}'|G^{00}(t'^{2}+\frac{G^{ij}}{G^{00}}V_{i}V_{j}t'^{2})
|{\cal P}'\rangle =\\ 
\langle{\cal P}'|G^{00}(t'^{2}-|v|_{q}^{2}t'^{2})|{\cal P}'\rangle&=&
G^{00}(t'^{2}-|v|_{q}^{2}t'^{2})=\\
\langle v_{3},|v|_{q}^{2}|v_{3},|v|_{q}^{2}\rangle\otimes
\langle{\cal P}_{0}|G^{IJ}X_{I}X_{J}|{\cal P}_{0}\rangle &=&
\langle v_{3},|v|_{q}^{2}|v_{3},|v|_{q}^{2}\rangle\otimes
\langle{\cal P}_{0}|G^{00}t^{2}|{\cal P}_{0}\rangle = \\
I&\otimes& G^{00}t^{2}
\end{eqnarray*}
leading to the $q$-deformed lifetime dilatation in the noncommutative special
relativity
\begin{eqnarray}
t'=\frac{t}{(1-|v|_{q}^{2})^{\frac{1}{2}}}.
\end{eqnarray}

In the semi classical limit $q\simeq 1\pm \kappa + O(\kappa^{2})$, where
$|v|_{q=0}=|v_{cl}|$ denotes the classical velocity, the correction to the
classical formula of the lifetime dilatation of unstable particles
$\Delta(t_{cl})$ is given by
\begin{eqnarray}
\Delta(t'_{q}) \simeq \Delta(t_{cl}) \mp \kappa\Delta(t_{cl})\frac{|v_{cl}|^{2}}
{1-|v_{cl}|^{2}} + O(\kappa^{2}).
\end{eqnarray}
Note that we can also consider the bicovariant bimodule ${\cal M}$ over
${\cal L}$ where, instead of space-time coordinates $X_{N}$, we take as basis
of vector space of all the right invariant elements of ${\cal M}$ the
coordinates $P_{N}$ of the energy-momentum four-vector. $P_{0}$ and $P_{n}$
are identified to the energy operator and the coordinates of the
vector-momentum operator respectively. In this case, we have only to replace
in this section  $X_{N}$ by $P_{N}$, $\tau^{2}$ by $G^{00}m^{2}$, the rest
state $|{\cal P}_{0}\rangle$ by $|m,0,m^{2}\rangle$ and the
$|{\cal P}'\rangle$ by $|E,p_{3},m^{2}\rangle$ where $E$ and $p_{3}$ are real eigenvalues of
$P'_{0}$ and $P'_{3}$ respectively and $m^{2}$ is the square of rest mass
which is bi-invariant, real and central. The mass is given by
$E^{2}-|p|_{q}^{2}=m^{2}$ where
$|p|_{q}^{2}=-\frac{G^{ij}}{G^{00}}P_{i}P_{j}$ is the length of the
vector-momentum. As for the space-time coordinates we can measure exactly and
simultaneously the mass, the energy, the length of the vector-momentum and
only one of its components, for instance $p_{3}$. From momentum version of
(30) we may see that the eigenvalues $E$ and $p_{3}$ are given in terms of
the mass and the velocity  by
\begin{eqnarray}
E=\frac{m}{(1-|v|_{q}^{2})^{\frac{1}{2}}}=m\gamma~~,~~
p_{3}=\frac{mv_{3}}{(1-|v|_{q}^{2})^{\frac{1}{2}}}=mv_{3}\gamma~~and~~
|p|_{q}=\frac{m|v|_{q}}{(1-|v|_{q}^{2})^{\frac{1}{2}}}
\end{eqnarray}
which are the quantum analog of the usual relativistic formulas of the
energy and the vector-momentum of free particles moving in the Minkowski
space with velocity $v$.
\section{Evolution of particles in quantum Minkowski space-time}
Having associated the quantum mechanics principles with the properties of the
quantum coordinates of the Minkowski space-time and their transformations under
the quantum Lorentz group, we investigated in this section the states
describing the evolution of particles in the quantum the Minkowski space-time.
Let $|t,x_{3},\tau^{2}\rangle$ be a state satisfying (21). From (10-11), we
see that the states $|n,t,x_{3},\tau^{2}\rangle$ and
$|-m,t,x_{3},\tau^{2}\rangle$ given by
\begin{eqnarray}
|n,t,x_{3},\tau^{2}\rangle=\overbrace{Z...Z}^{n}|t,x_{3},\tau^{2}\rangle
~and~ |-m,t,x_{3},\tau^{2}\rangle=\overbrace{\overline{Z}...\overline{Z}}^{m}
|t,x_{3},\tau^{2}\rangle
\end{eqnarray}
are eigenstates of $X_{3}$ with eingenvalues given respectively by
\begin{eqnarray}
X_{3}|n,t,x_{3},\tau^{2}\rangle = (q^{2n}x_{3} + q^{-1}(q^{2n}-1)t)
|n,t,x_{3},\tau^{2}\rangle=x_{3}^{(n)}|n,t,x_{3},\tau^{2}\rangle,\\
X_{3}|-m,t,x_{3},\tau^{2}\rangle = (q^{-2m}x_{3} + q^{-1}(q^{-2m}-1)t)
|-m,t,x_{3},\tau^{2}\rangle=x_{3}^{(-m)}|-m,t,x_{3},\tau^{2}\rangle.
\end{eqnarray}
In following we shall assume that $q>0$. Therefore we can see that these
eigenvalues satisfy
\begin{eqnarray}
x_{3}^{(n)}\leq x_{3} ~and~ x_{3}^{(-n)}\geq x_{3}~~~~~for~0<q<1,\\
x_{3}^{(n)}\geq x_{3} ~and~ x_{3}^{(-n)}\leq x_{3}~~~~~~~~for~q>1
\end{eqnarray}
then for $0 <q< 1~(q > 1)$, $Z$ and $\overline{Z}$ are lowering (raising)
and raising (lowering) operators for the engenvalues of $X_{3}$. To establish
the conditions on the eigenvalues $x_{3}$, we consider the relations
\begin{eqnarray}
(a)~~Z\overline{Z}= q^{-2}(X_{0}+qX_{3})(X_{0}-q^{-1}X_{3})-
q^{-\frac{1}{2}}\tau^{2},\nonumber\\
(b)~~~~\overline{Z}Z= q^{-2}(X_{0}+qX_{3})(X_{0}-q^{3}X_{3})-q^{-\frac{1}{2}}\tau^{2}
\end{eqnarray}
obtained from (13) and the commutation relation (12). Since the states must
be of positive norm, the main value of (39) satify:
\begin{eqnarray}
(a)~~q^{-2}(t+qx_{3})(t-q^{-1}x_{3})-q^{-2}\alpha^{2}\geq 0,\nonumber\\
(b)~~~~q^{-2}(t+qx_{3})(t-q^{3}x_{3})-q^{-2}\alpha^{2}\geq 0
\end{eqnarray}
where $\tau^{2}=q^{-\frac{3}{2}}\alpha^{2}$. From (28), we can replace $X_{i}$
by $V_{i}t$ into the norm of the vector $X_{I}$ to get
$q^{-\frac{3}{2}}t^{2} - G^{ij}V_{i}V_{j}t^{2} =q^{-\frac{3}{2}}\alpha^{2}$
or $t^{2}(1-|v|_{q}^{2})=\alpha^{2}$, hence
$\alpha^{2}=\frac{t^{2}}{\gamma^{2}}$.
The conditions (40$a$-$b$) define the interval of the eigenvalues $x_{3}$ in
which it is possible to construct states of positive norm belonging to Hilbert
space. To define this interval, we begin to give the roots of (40$a$-$b$)
\begin{eqnarray}
x_{03}^{(a)\pm} =\frac{(q-q^{-1})\pm
(Q^{2}-\frac{4}{\gamma^{2}})^{\frac{1}{2}}}{2}t~~~~~~~~~where~~
x_{03}^{(a)-}<x_{03}^{(a)+},\\
x_{03}^{(b)\pm} =-q^{-2}\frac{(q-q^{-1})\pm
(Q^{2}-\frac{4}{\gamma^{2}})^{\frac{1}{2}}}{2}t~~~~~~where~~
x_{03}^{(b)+}<x_{03}^{(b)-}.
\end{eqnarray}
The reality of $x_{3}$ requires $Q^{2}-\frac{4}{\gamma^{2}}\geq 0$,
hence $\gamma^{2}\geq\frac{4}{Q^{2}}$. It is not very difficult to see
that $x_{03}^{(a)-}<x_{03}^{(b)+}$ and $x_{03}^{(a)+}<x_{03}^{(b)-}$ if
$0<q< 1$ and $x_{03}^{(b)+}<x_{03}^{(a)-}$ and
$x_{03}^{(b)-}<x_{03}^{(a)+}$ if $q > 1$. So the conditions (40$a$-$b$) are
both satisfied in the interval
$(x_{03}^{(a)-}<x_{03}^{(a)+})\cap (x_{03}^{(b)+}<x_{03}^{(b)-})$ which must
be nonempty. This requires that
\begin{eqnarray}
(a')~~~x_{03}^{(b)+}<x_{03}^{(a)+}~~~for~0<q< 1~and~~(b')~~
x_{03}^{(a)-}<x_{03}^{(b)-}~~~for~q > 1.
\end{eqnarray}
The relation (43$b'$) is satisfyied
if $(q-q^{-1}) - (Q^{2}-\frac{4}{\gamma^{2}})^{\frac{1}{2}}\leq -
q^{-2}(q-q^{-1}) + q^{-2}(Q^{2}-\frac{4}{\gamma^{2}})^{\frac{1}{2}}
\Rightarrow (1-q^{-2})Q\leq
(1+q^{-2})(Q^{2}-\frac{4}{\gamma^{2}})^{\frac{1}{2}}$ or
$(1-q{-2})^{2}Q^{2}\leq(1+q^{-2})^{2}(Q^{2}-\frac{4}{\gamma^{2}})$
implying
\begin{eqnarray}
\gamma^{2} \geq \frac{q^{2}(1+q^{-2})^{2}}{Q^{2}} = 1
\end{eqnarray}
the relation (43$a'$) gives the same condition (44) on $\gamma$. This
relation is necessary for the existence of Hilbert states of positive norm
(physical states), hence the causality principle in noncommutative special
relativity which states that there exist physical states only for real
velocities in the range $0\leq |v|_{q} \leq c = 1$.\\
Now, suppose that $x_{3} \in (x_{03}^{(b)+},x_{03}^{(a)+})$, $(0<q<1)$, is an
eigenvalue of the eigenstate $|t,x_{3},\tau^{2}\rangle$ of $X_{3}$. If we
consider the state $|-m,t,x_{3},\tau^{2}\rangle$ eigenvector of $X_{3}$ with
eigenvalue $q^{-2m}x_{3} + q^{-1}(q^{-2m}-1)t > x_{3}$ such that
$q^{-2(m+1)}x_{3} + q^{-1}(q^{-2(m+1)}-1)t > x_{03}^{(a)+}$, since the state
norms are positive, then
\begin{eqnarray*}
q^{-2m}x_{3} + q^{-1}(q^{-2m}-1)t = x_{03}^{(a)+} = \frac{(q-q^{-1})}{2}t +
\frac{(Q^{2} -\frac{4}{\gamma^{2}})^{\frac{1}{2}}}{2}t
\end{eqnarray*}
implying
\begin{eqnarray}
x_{3} = q^{2m}(\frac{Q}{2} +
\frac{(Q^{2} -\frac{4}{\gamma^{2}})^{\frac{1}{2}}}{2})t -q^{-1}t.
\end{eqnarray}
Since the operator $Z$ decreases the value of $x_{3}$, it exists a number
$r\in N^{+}$ such that $\overbrace{Z...Z}^{r}
|t,x_{3},\tau^{2}\rangle$ is eigenstate of $X_{3}$ with eigenvalue
$x_{03}^{(b)+}$, hence
\begin{eqnarray}
q^{2r}x_{3} + q^{-1}(q^{2r} -1)t= -q^{-2}(\frac{(q-q^{-1})}{2}+
\frac{(Q^{2} -\frac{4}{\gamma^{2}})^{\frac{1}{2}}}{2})t.
\end{eqnarray}
By setting (45) into (46), we get
\begin{eqnarray}
q^{2(r+m)}(\frac{Q}{2} +
\frac{(Q^{2} -\frac{4}{\gamma^{2}})^{\frac{1}{2}}}{2}) =q^{-2}(\frac{Q}{2}
-\frac{(Q^{2} -\frac{4}{\gamma^{2}})^{\frac{1}{2}}}{2})
\end{eqnarray}
from which we obtain
\begin{eqnarray*}
\frac{Q^{2}}{4}(q^{2(r+m+1)} - 1)^{2} =
\frac{Q^{2} -\frac{4}{\gamma^{2}}}{4}(q^{2(P+m+1)} + 1)^{2}
\end{eqnarray*}
leading to
\begin{eqnarray}
\gamma^{2} = \frac{(q^{(r+m+1)} +q^{-(r+m+1)})^{2}}{Q^{2}}.
\end{eqnarray}
On the other hand, if we consider the state $|n,t,x_{3},\tau^{2}\rangle$
eigenvector of $X_{3}$ with eigenvalue $q^{2n}x_{3} + q^{-1}(q^{2n}-1)t <
x_{3}$ such that $q^{2(n+1)}x_{3} + q^{-1}(q^{2(n+1)}-1)t < x_{03}^{(b)-}$, we
have necessary
\begin{eqnarray*}
q^{2n}x_{3} + q^{-1}(q^{2n}-1)t = x_{03}^{(b)-} =-q^{-2}
(\frac{(q-q^{-1})}{2} +
\frac{(Q^{2} -\frac{4}{\gamma^{2}})^{\frac{1}{2}}}{2})t
\end{eqnarray*}
implying
\begin{eqnarray}
x_{3}=q^{-2(n+1)}(\frac{Q}{2} -
\frac{(Q^{2} -\frac{4}{\gamma^{2}})^{\frac{1}{2}}}{2})t - q^{-1}t.
\end{eqnarray}
Since the operator $\overline{Z}$ increases the value of $x_{3}$, there exists
a number $s\in N^{+}$ such that $\overbrace{\overline{Z}...\overline{Z}}^{s}
|t,x_{3},\tau^{2}\rangle$ is a eigenstate of $X_{3}$ with eigenvalue
$x_{03}^{(a)+}$ yielding
\begin{eqnarray}
q^{-2s}x_{3} + q^{-1}(q^{-2s} -1)t= \frac{(q-q^{-1})}{2}t+
\frac{(Q^{2} -\frac{4}{\gamma^{2}})^{\frac{1}{2}}}{2}t.
\end{eqnarray}
By setting (49) into (50), we get
\begin{eqnarray*}
q^{-2(s+n+1)}(\frac{Q}{2} -
\frac{(Q^{2} -\frac{4}{\gamma^{2}})^{\frac{1}{2}}}{2}) =\frac{Q}{2}
+\frac{(Q^{2} -\frac{4}{\gamma^{2}})^{\frac{1}{2}}}{2}
\end{eqnarray*}
which is equivalent to (47). Therefore $\gamma^{2}$ is quantized and given
by
\begin{eqnarray}
(\gamma^{(L)})^{2} = \frac{(q^{(s+n+1)} +q^{-(s+n+1)})^{2}}{Q^{2}}
= \frac{(q^{(r+m+1)} +q^{-(r+m+1)})^{2}}{Q^{2}}
= \frac{(q^{(L+1)} +q^{-(L+1)})^{2}}{Q^{2}}
\end{eqnarray}
where $L=0,1,2,..., \infty$. By substituting (51) into (45), we get the states
$|t,x_{3}^{(L,n)},\tau^{2}\rangle$ such that
$X_{3}|t,x_{3}^{(L,n)},\tau^{2}\rangle =x_{3}^{(L,n)}
|t,x_{3}^{(L,n)},\tau^{2}\rangle$ where
\begin{eqnarray}
x_{3}^{(L,n)}= (Q\frac{q^{-(L+1-2n)}}{q^{(L+1)}+q^{-(L+1)}} - q^{-1})t
\end{eqnarray}
and $n=0,1, ...,L$. Note that if we set $L=2l$ and $m=l-n$ we can rewrite the
states as $|t,x_{3}^{(l,m)},\tau^{2}\rangle$ eigenstates of $X_{3}$ with eigenvalues
\begin{eqnarray}
x_{3}^{(l,m)}=q^{-1}(Q\frac{q^{2m}}{q^{(2l+1)}+q^{-(2l+1)}} - 1)t =
q^{-1}(\frac{q^{2m}}{\gamma^{(l)}} - 1)t
\end{eqnarray}
where $\gamma^{(l)} = \frac{q^{(2l+1)} +q^{-(2l+1)}}{Q}$, with
$l=0,\frac{1}{2},1,....\infty$ and $m$ runs by integer steps over the
range $-l\leq m \leq l$. The substitution of (51) into (49) gives the same
states with the same eigenvalues (52). Following the same analysis as above,
we can show that the case $q > 1$ gives the same results for $\gamma$
(51) and for $x_{3}$ (52). From (30) and (53) we deduce the quantization of
the velocity as
\begin{eqnarray}
|v|_{q}^{2}=1-\frac{1}{(\gamma^{(l)})^{2}}~~~and~~~
v_{3}^{(l,m)}=q^{-1}(\frac{q^{2m}}{\gamma^{(l)}} - 1).
\end{eqnarray}
Note that in the case where we consider the energy-momentum four-vector
$P_{N}$ instead of the coordinates $X_{N}$. The relation (33) combined with
(51) and (54) show that the energy $E$ and the component $P_{3}$ of the
vector-momentum present discrete spectrums given by $E^{(l)}=m\gamma^{(l)}$
and $P_{3}^{(l,m)}=mv_{3}^{(l,m)}\gamma^{(l)}$. The basis of the Hilbert
space states of free particle in the momentum representation is given by
$|E^{(l)},p_{3}^{(l,n)},m\rangle$ normalized as
$\langle E^{(l)},p_{3}^{(l,n)},m|E^{(l')},p_{3}^{(l',n')},m\rangle =
\delta_{l,l'}\delta_{n,n'}$. where $l,l'=0,\frac{1}{2},1,....\infty$ and
$-l\leq m,m' \leq l$. 
\section{Discussions and conclusions}
In this paper, we have seen that an adequat association of the quantum
mechanics principles with the properties of the quantum Minkowski space-time
and its quantum Lorentz transformations make possible the description of the
evolution of particles in the noncommutative special relativity. In this
formalism we have showed that only the length of the velocity and one of its
components can be measured exactly and present discret spectrums. From the
quantum boost, we have established the principle of causality and the
$q$-deformed analog of the lifetime dilatation formula for moving unstable
particles from which we see that
$\frac{\Delta(t'_{q})}{\Delta(t_{q})} =\gamma^{(l)}$,
($l,=0,\frac{1}{2},1,....\infty$) is quantized. This quantization presents an
interesting novelty because as opposite to the costumary believe where we
consider that the effects of noncommutativity of space may be observed only
at very high energy, Planck scale $M_{P}$, which is beyond the reach of
conceivable experiments, in this scenario, the effects of the evolution of
free particles in the noncommutative special relativity are expressed in terms
of quantized velocities (hence quantized currents for charged particles) which
require not very energies but high precision on measurements.\\
\vskip0.1truecm
Note that the case $L=0,~n=0$ corresponds to $x_{3}=0$ and
$\gamma_{q}^{2} =1$ yielding $|v|=0$ and corresponds to the unique state
describing a particle at rest considered in the section 3.\\
\vskip0.1truecm
In the case where $\tau^{2}=0$, corresponding to a particle on the light cone,
the roots of (40$a$-$b$) reduce to $x_{03}^{(a)-}=-q^{-1}t<x_{03}^{(a)+}=qt$ and
$x_{03}^{(b)+}=-q^{-1}t<x_{03}^{(b)-}=q^{-3}t$. Then the conditions (40$a$-$b$)
are both satisfied in the interval
$(x_{03}^{(a)-}=x_{03}^{(b)+},x_{03}^{(a)+})$ if $0<q<1$ or in the interval
($x_{03}^{(a)-}=x_{03}^{(b)+},x_{03}^{(b)-})$ if $q >1$.The states of positive norm are given by
$|t,x_{3}^{(n)},\tau^{2}\rangle$ where the engenvalues of $X_{3}$ are given
by
\begin{eqnarray}
x_{3}^{(n)} = q^{2n}Qt-q^{-1}t,~~~~~~for~0<q\leq 1\\
x_{3}^{(n)} = q^{-2(n+1)}Qt-q^{-1}t,~~~~~for~q\geq 1
\end{eqnarray}
with $n=0,1,..., \infty$. Note that the eigenstate $|t,-q^{-1}t,0\rangle$ is
stable under the action of $Z$ and $\overline{Z}$ in the sens that
$\forall n,m\in N^{+},~
\overbrace{Z...Z}^{n}\overbrace{\overline{Z}...\overline{Z}}^{m}
|t,-q^{-1}t,0\rangle$ is an eigenstate of $X_{3}$ with eigenvalue $-q^{-1}t$
hence $v_{3}=-q^{-1}$.
$\langle t,-q^{-1}t,0|Z\overline{Z}|t,-q^{-1}t,0\rangle$ and
$\langle t,-q^{-1}t,0|\overline{Z} Z|t,-q^{-1}t,0\rangle $ vanish, therefore,
the light velocity reduce to the $|v|_{q}^{2}=q^{2}v_{3}^{2}=1$ which is,
by virtue of (31), the upper limit of velocities. We may retrieve this case
in the limit $L \rightarrow \infty\Rightarrow \tau^{2}
\rightarrow 0$.\\
\vskip0.1truecm
To boost again the state $|t,x_{3},\tau^{2}\rangle$ we must know explicitly
all the commutation relations of the sixteen generators $\Lambda_{N}^{~M}$ to
get the set of commuting generators and their eigenstate $|sym_{q}\rangle$
labelled by its different eigenvalues. This study permits to define the
addition rules of velocities in the noncommutative special relativity [8].
\vskip0.1truecm
When we restrict the generators of the quantum $SL(2,C)$ group to those of
the $SU(2)$ by imposing unitarity condition, we get relations (5) and (6)
which lead us to the restriction of the Minkowski space-time transformations
to the orthogonal transformations of the three dimensional space $R_{3}$
equipped with the coordinate system $X_{i}$, ($i=1,2,3$). These
transformations leave invariant the time coordinate $X_{0}$. In fact, as a
consequence of (5) and (6) we have
\begin{eqnarray}
\Delta_{L}(X_{0}) = \overline{\Lambda}_{0}^{~0}\otimes X_{0}
= I\otimes X_{0}\nonumber\\
\Delta_{L}(X_{i}) = \overline{\Lambda}_{i}^{~j}\otimes X_{j}
\end{eqnarray}
where the generators $\overline{\Lambda}_{i}^{~j}=
\frac{1}{Q}
\overline{\sigma}_{i\dot{\gamma}}^{~~\alpha}M_{\alpha}^{~\sigma}
\sigma^{j}_{~\sigma\dot{\rho}}M_{\dot{\beta}}^{~\dot{\rho}}
\varepsilon^{\dot{\gamma}\dot{\beta}}=\frac{1}{Q}
\overline{\sigma}_{i\dot{\gamma}}^{~~\alpha}M_{\alpha}^{~\sigma}
\sigma^{j}_{~\sigma\dot{\rho}}S(M_{\rho}^{~\beta})
\varepsilon^{\dot{\gamma}\dot{\beta}}$ generate a Hopf sub-algebra
${\cal SO}_{q}(3)$ of $\cal L$ whose the axiomatic structure is derived from
those of $\cal L$ as $\Delta(\overline{\Lambda}_{i}^{~j}) =
\overline{\Lambda}_{i}^{~k}\otimes \overline{\Lambda}_{k}^{~j}$,
$\varepsilon(\overline{\Lambda}_{i}^{~j})= \delta_{i}^{~j}$ and
$S(\overline{\Lambda}_{i}^{~j}) = G_{iK}
\overline{\Lambda}_{L}^{~K}G^{~Lj} = G_{ik}
\overline{\Lambda}_{l}^{~k}G^{~lj}$ where
$G^{~ij}$ is the restriction of $G^{~IJ}$ to the quantum space $R_{3}$
satisfying $G^{~ik}G_{kj} = \delta_{j}^{i} =G_{jk}G^{~ki}$.
The form of the antipode of $\overline{\Lambda}_{i}^{~j}$ implies
the orthogonality properties
\begin{eqnarray*}
G^{~ij} \overline{\Lambda}_{i}^{~l}\overline{\Lambda}_{j}^{~k}
= G^{~lk} &~and&
G_{lk} \overline{\Lambda}_{i}^{~l} \overline{\Lambda}_{i}^{~k}= G_{ij}.
\end{eqnarray*}
Therefore, $\overline{\Lambda}_{i}^{~j}=\frac{1}{Q}
\overline{\sigma}_{i\dot{\gamma}}^{~~\alpha}M_{\alpha}^{~\sigma}
\sigma^{j}_{~\sigma\dot{\rho}}S(M_{\rho}^{~\beta})
\varepsilon^{\dot{\gamma}\dot{\beta}}$ establishes a correspondence between
$SU_{q}(2)$ and $SO_{q}(3)$ group. In fact the two-dimensional
representation of $SU_{q}(2)$ is given by $\left(\begin{array}{cc}
\alpha& q\gamma^{\star}\\-\gamma&\alpha^{\star} 
\end{array}
\right)$[9] and in the three dimensional space spanned by the basis
$Qe_{-1}=Z$, $e_{0} =X_{3}$ and $Qe_{1}=\overline{Z}$, the generators
$\overline{\Lambda}_{i}^{~j}=\frac{1}{Q}
\overline{\sigma}_{i\dot{\gamma}}^{~~\alpha}M_{\alpha}^{~\sigma}
\sigma^{j}_{~\sigma\dot{\rho}}S(M_{\rho}^{~\beta})
\varepsilon^{\dot{\gamma}\dot{\beta}}$ write
\begin{eqnarray*}
(d_{1,ij})_{i,j=-1,0,1}=
\left(\begin{array}{clcr}
\alpha^{\star 2} & -(1+q^{2})\alpha^{\star}\gamma &-q\gamma^{2}\\
\gamma^{\star}\alpha^{\star} & 1-(1+q^{2})\gamma^{\star}\gamma & \alpha\gamma\\
-q\gamma^{\star 2}& -(1+q^{2})\gamma^{\star}\alpha & \alpha^{2}
\end{array}
\right)\in M_{3}\otimes C(SU_{q}(2))
\end{eqnarray*}
which is an irreducible three-dimensional representation of $SU_{q}(2)$
considered in [9]. In the other hand if we replace
$\overline{Z}=Qe_{1}$, $X_{3}=e_{0}$,
$Z=Qe_{-1}$, $\lambda = (q-q^{-1})t=(q-q^{-1})X_{0}$ and
$\rho =q^{-2}t^{2} - q^{-\frac{1}{2}}\tau^{2}$ into (19-22) we retrieve the
algebra of the quantum spheres given by (2$a$-$e$) of [10]. In the framework
developped above, the states $|t,x_{3}^{(n)},\tau^{2}\rangle$ whose
eigenvalues are given by (55-56) (case $\tau^{2}=0$) represent a basis of the
Hilbert space where the quantum sphere algebra is represented. The fixed time
$t$ represents the ray of the sphere.\\
\vskip0.1truecm
{\bf Acknowledgments.} I am grateful to M. Dubois-Violette for his kind
interest and hepful suggestions. I am grateful for hospitatility at the Abdus
Salam International centre for theoretical physics where the extended version
of this work was done.\\
{\bf References:}
1)U. Meyer, Commun. Math. Phys. 174(1996)457. P. Podle\'s, Commun. Math. Phys.
181(1996)569.\\
2)M. Pillin, W. B. Schmidke and J. Wess, Nucl. Phys. B(1993)223. M. Pillin
and Weilk, J. Phys. A: Math. Gen. 27(1994)5525. B. L. Cerchiai and J. Wess
:"q-Deformed Minkowski Space based on q-Lorentz Algebra" LMU-TPW 98-02,
MPI-PhT/89-09.\\
3)M. Lagraa, "On the quantum Lorentz group", to appear in J. Geom. Phys.\\
4)M. Lagraa, Int. J. Mod. Phys. A11(1996)699.\\
5)M. Dubois-Violette and G. Launer, Phys. Lett. B245(1990)175.\\
6)O. Ogiesvestsky, W. B. Schmidke, J. Wess and B. Zumino, Commun. Math.
Phys. 150(1992)495.\\
7)U. Carow-Watamura, M. Schlieker, M. Scholl and S. Watamura, Z. Phys. C-
$particles~and~fields$ 48(1990)159, Int. J. Mod. Phys. A6(1991)3081.\\
8)M. Lagraa, work in progress.\\
9)S. L. Woronowicz, Publ. RIMS, Kyoto Univ. 23 (1987) 117, Commun. Math.
Phys. 122(1989)125.\\
10)P. Podle\'s, Lett. Math. Phys. 14(1987)193, Lett. Math. Phys. 18(1989)107,
Commun. Math. Phys. 170(1995)1.\\
\end{document}